\documentclass[reprint,superscriptaddress,showpacs,amsmath,amssymb,twocolumn,prb]{revtex4-1}

\usepackage{graphicx}% Include figure files
\usepackage{dcolumn}% Align table columns on decimal point
\usepackage{bm}% bold math
\usepackage{verbatim}% for multiline comments
\usepackage[usenames, dvipsnames]{xcolor}
\usepackage{float}
\usepackage{upgreek}

\def\be{\begin{equation}}
\def\ee{\end{equation}}
\def\bea{\begin{eqnarray}}
\def\eea{\end{eqnarray}}

\begin{document}

% \title{Coherent properties of localized excitons and trions in ZnO/(Zn,Mg)O quantum wells}

\title{Coherent dynamics of localized excitons and trions in ZnO/(Zn,Mg)O quantum wells studied by photon echoes}

\author{I.~A.~Solovev}
\affiliation{Experimentelle Physik 2, Technische Universit\"at Dortmund, 44221 Dortmund, Germany}
\affiliation{Spin Optics Laboratory, St.~Petersburg State University, 198504 St.~Petersburg, Russia}
\author{S.~V.~Poltavtsev}
\email{sergei.poltavtcev@tu-dortmund.de}
\affiliation{Experimentelle Physik 2, Technische Universit\"at Dortmund, 44221 Dortmund, Germany}
\affiliation{Spin Optics Laboratory, St.~Petersburg State University, 198504 St.~Petersburg, Russia}
\author{Yu.~V.~Kapitonov}
\affiliation{Physics Faculty, St.~Petersburg State University, 199034 St.~Petersburg, Russia}
\author{I.~A.~Akimov}
\affiliation{Experimentelle Physik 2, Technische Universit\"at Dortmund, 44221 Dortmund, Germany}
\affiliation{Ioffe Physical-Technical Institute, Russian Academy of Sciences, 194021 St.~Petersburg, Russia}
\author{S.~Sadofev}
\affiliation{AG Photonik, Institut f\"ur Physik, Humboldt-Universit\"at zu Berlin, D-12489 Berlin, Germany}
\author{J.~Puls}
\affiliation{AG Photonik, Institut f\"ur Physik, Humboldt-Universit\"at zu Berlin, D-12489 Berlin, Germany}
\author{D.~R.~Yakovlev}
\affiliation{Experimentelle Physik 2, Technische Universit\"at Dortmund, 44221 Dortmund, Germany}
\affiliation{Ioffe Physical-Technical Institute, Russian Academy of Sciences, 194021 St.~Petersburg, Russia}
\author{M.~Bayer}
\affiliation{Experimentelle Physik 2, Technische Universit\"at Dortmund, 44221 Dortmund, Germany}
\affiliation{Ioffe Physical-Technical Institute, Russian Academy of Sciences, 194021 St.~Petersburg, Russia}

\date{\today}

\begin{abstract}
We study optically the coherent evolution of trions and excitons in a $\updelta$-doped 3.5 nm-thick ZnO/Zn$_{0.91}$Mg$_{0.09}$O multiple quantum well by means of time-resolved four-wave mixing at temperature of 1.5~K. Employing spectrally narrow picosecond laser pulses in the $\upchi^{(3)}$ regime allows us to address differently localized trion and exciton states, thereby avoiding many-body interactions and excitation-induced dephasing. The signal in the form of photon echoes from the negatively charged A excitons (T$_\text{A}$, trions) decays with coherence times varying from 8 up to 60~ps, depending on the trion energy: more strongly localized trions reveal longer coherence dynamics. The localized neutral excitons decay on the picosecond timescale with coherence times up to $T_2=4.5$~ps. The coherent dynamics of the X$_\text{B}$ exciton and T$_\text{B}$ trion are very short ($T_2<1$~ps), which is attributed to the fast energy relaxation from the trion and exciton B states to the respective A states. The trion population dynamics is characterized by the decay time $T_1$, rising from 30~ps to 100~ps with decreasing trion energy. 
\end{abstract}

\maketitle

\section{Introduction}
\label{sec:1}

Zinc-oxide is a remarkable semiconductor combining many appealing properties in one material \cite{KlingshirnReview2007}. Its large exciton binding energy of 60~meV makes ZnO important for numerous optical applications, including optically pumped  ultraviolet lasers, also with vertical cavity \cite{BagnallAPL1997,SadofevAPL2011}, light emitting diodes \cite{KawasakiNM2004}, solar cells \cite{Yip2008}, spintronics devices \cite{BratschitschNL2011}, and others\cite{Chen2010}. Among the variety of ZnO-based materials, epitaxial ZnO quantum wells (QW) are of great interest, because the exciton properties can be fine-tuned by adjusting structural parameters such as the QW thickness\cite{GilPRB2005} or the barrier composition\cite{GilAPL2007} to match the needs of the targeted applications. The optical properties of the QWs include also different exciton complexes, in particular, negatively charged excitons (trions, T) formed from resident electrons provided by  doping. Trions were observed in ZnO/(Zn,Mg)O QWs and studied by different magneto-optical methods\cite{MakinoPRB2009, PulsPRB2012}. To explore the coherent properties of excitons and exciton complexes, the methods of four-wave-mixing and photon echoes are very efficient \cite{BermanMalinovskyBook}. They were applied to study the picosecond coherent dynamics of donor-bound excitons in epitaxial films of ZnO at low temperatures \cite{PoltavtsevZnO2017}, the homogeneous and inhomogeneous exciton broadenings in bulk ZnO\cite{ShubinaPRB2013}, and the coherent biexciton dynamics in ZnO/(Zn,Mg)O QWs at room temperature \cite{DavisAPL2006}. It was found, in particular, that the bulk excitons in ZnO show a sub-ps coherent dynamics even at low temperatures\cite{PoltavtsevZnO2017, ShubinaPRB2013}. The donor-bound A~excitons (D$^0$X$_\text{A}$) in ZnO epitaxial layers, conversely, demonstrate much longer dephasing times reaching 60~ps, while corresponding D$^0$X$_\text{B}$ states show substantially shorter times of about 3.5~ps\cite{PoltavtsevZnO2017}. The coherent optical properties of the trions, which are important in the context of coherent manipulation of either optical excitations or spin states in ZnO, have not been addressed so far.

In this paper, we explore the coherent dynamics of trions and excitons in ZnO/(Zn,Mg)O quantum wells by means of the time-resolved four-wave-mixing (FWM) technique. Using spectrally-narrow picosecond laser pulses we individually address the trion and exciton states and detect the coherent response in the form of photon echoes. We find that the T$_\text{A}$ trions demonstrate a long-lived coherent evolution reaching the tens-of-picosecond timescale. Also, the coherence and population relaxation times of T$_\text{A}$ strongly depend on its localization degree on the quantization potential fluctuations. The B exciton and trion, however, demonstrate strongly damped echoes revealing a major difference in the coherent behavior compared to the A exciton and trion, respectively.

\begin{figure}[t]
	\vspace{5mm}
	\includegraphics[width=\linewidth]{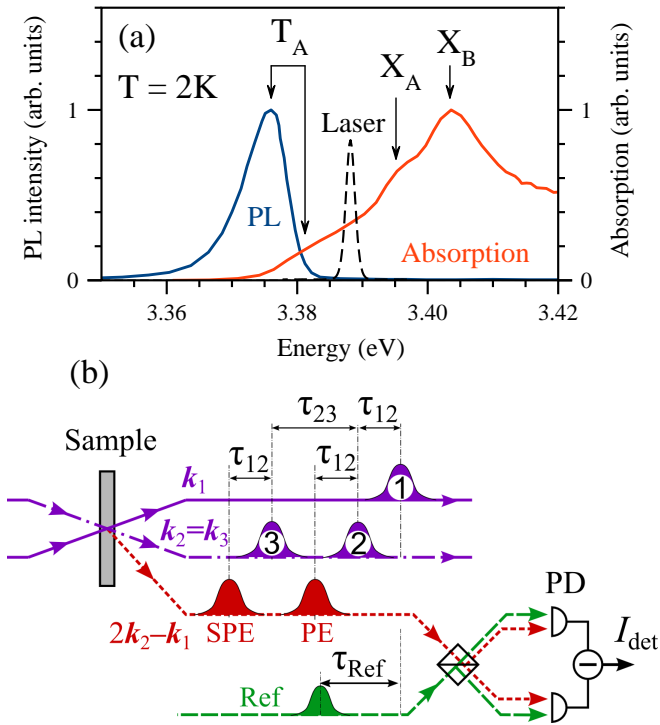}
	\caption{(a) PL and absorption spectra of the ZnO/(Zn,Mg)O MQW measured at $T=2$~K. Features related to the T$_\text{A}$ trion and X$_\text{A}$, X$_\text{B}$ excitons are indicated with arrows. The spectrum of the laser pulse used in the FWM experiments with full-width at half-maximum of 1.7~meV is shown by the dashed line. (b) Scheme of optical excitation of the sample and detection of the time-resolved four-wave mixing in the transmission geometry. PD stands for the photodetector. }
	\label{setup}
\end{figure}

\section{Sample and Experimental Method}
\label{sec:2}

The studied ZnO/(Zn,Mg)O multiple QW (MQW) sample was grown by radical-source molecular-beam epitaxy on a $c$-plane sapphire wafer. The sample consists of a 2-nm-thick MgO nucleation layer, a 500-nm-thick (Zn,Mg)O buffer, five 3.5~nm-thick ZnO QWs separated by 8.5~nm-thick (Zn,Mg)O barriers, and a (Zn,Mg)O cap layer. To provide resident electrons the MQW part was doped by Sb $\delta$-layers \cite{LiuJAP2012}, symmetrically placed inside the barriers and at a distance of 4~nm before the first QW and behind the top one. Since ZnO is intrinsically n-type, in QWs without doping, resident electrons lead to trion transitions in optical spectra\cite{PulsPRB2012}. The delta-doping used here is to achieve higher electron densities and therefore more intense trion transitions for stronger FWM signals. The sample was grown at a wafer temperature of $300^\circ$C. Before deposition of the MQW part the buffer layer was additionally annealed during 5 minutes at $590^\circ$C. The Mg content ($x$) in the buffer and the barriers amounts to $x=0.09$, which leads to small built-in electric fields\cite{GilAPL2007}.

Photoluminescence (PL) and absorption spectra of the studied MQW measured using the same procedures as in Ref.~\cite{PulsPRB2012} are plotted in Fig.~\ref{setup}(a). They display several features similar to those that were investigated recently in a ZnO/(Zn,Mg)O MQW by magneto-optical measurements \cite{PulsPRB2012,PulsPRB2016}. The optical transition located at energy 3.375~eV in the PL spectrum is identified as the T$_\text{A}$ trion. In the absorption spectrum it is manifested as a shoulder at approximately 3.381~eV. The estimated Stokes shift of about 6~meV is in accord with the 9~meV PL band halfwidth, signifying the dominant inhomogeneous broadening of the T$_\text{A}$ transitions due to localization. The absorption spectrum also contains two high-energy features, corresponding to the A and B excitons (X$_\text{A}$, X$_\text{B}$) located at 3.395 and 3.405~eV, respectively.

In order to access the coherent dynamics of these transitions we implement the four-wave-mixing technique with picosecond temporal resolution. The sample is immersed in liquid helium inside the variable temperature insert of a cryostat and cooled down to the temperature of $T=1.5$~K. Optical excitation of the sample with photon energy of about 3.4~eV ($\sim$365~nm) is performed by spectrally narrow laser pulses with duration of 1.3~ps, which are obtained from the output of a tunable picosecond Ti:sapphire mode-locked laser by second harmonic generation. The pulse spectrum is shown in Fig.~\ref{setup}(a), its bandwidth of about 1.7~meV is clearly smaller than the inhomogeneous broadening of the trion and exciton transitions. The laser pulse is split into three exciting pulses delayed relative to each other by the time intervals $\uptau_{12}$ and $\uptau_{23}$, as shown in Fig.~\ref{setup}(b). The first and second pulses hit the sample with $\bm{k}_1$ and $\bm{k}_2$ wavevectors, respectively, oriented close to the sample normal with an angle of about $1^\circ$ between each other, while the third pulse propagated the same path as the second one, $\bm{k}_3=\bm{k}_2$. Linearly co-polarized pulses are focused into a spot of about 300~$\mu$m in diameter. The FWM signal is collected in transmission geometry along the direction $2\bm{k}_2-\bm{k}_1$. Optical heterodyning and interferometric detection using an additional reference pulse (delayed by the time $\uptau_\text{Ref}$) are applied, as described in Ref.\cite{PoltavtsevZnO2017}, in order to accomplish a high-sensitivity background-free detection of the FWM amplitude. We checked that even application of the maximal available power density of the exciting pulses (200~nJ/cm$^2$) does not go beyond the $\upchi^{(3)}$ regime of the FWM measurements and does not affect significantly the measured coherence times.

\begin{figure}[t]
	\vspace{5mm}
	\includegraphics[width=\linewidth]{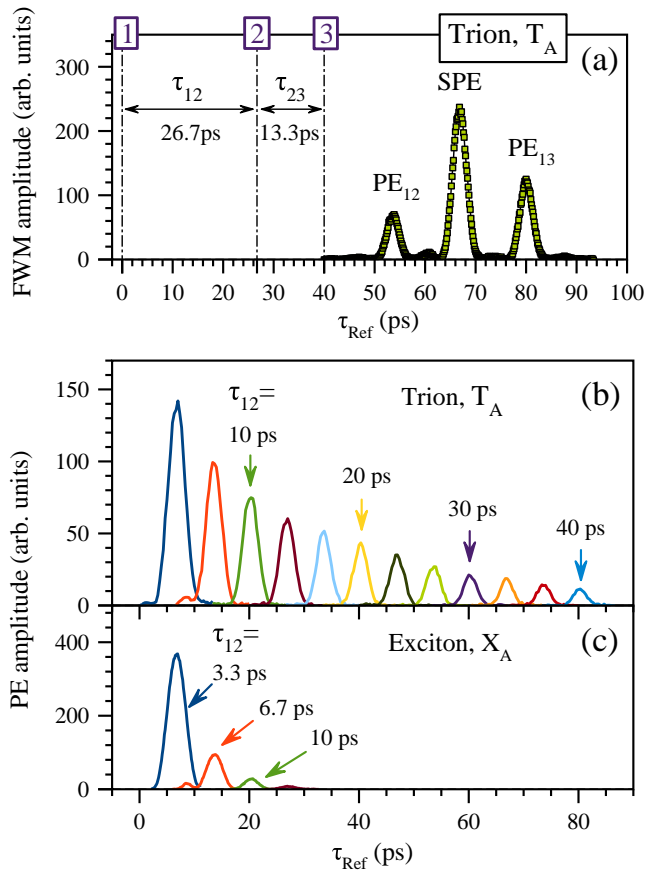}
	\caption{(a) FWM transient measured from the T$_\text{A}$ trion (3.3773~eV) in the three-pulse excitation regime at $T=1.5$~K: PE$_{12}$ and PE$_{13}$ are the two-pulse photon echoes resulting from the pulse doublets 1-2 and 1-3, respectively. SPE is the three-pulse photon echo. The PE$_{13}$ peak is higher than the PE$_{12}$ peak, because the intensity of the pulse 3 was four times stronger than that of the pulse 2. (b), (c) Time-resolved two-pulse photon echoes measured from the T$_\text{A}$ trion (3.378~eV) and X$_\text{A}$ exciton (3.394~eV), respectively, at various pulse delays $\tau_{12}$, indicated by the arrows.}
	\label{transients}
\end{figure}

\section{Experimental Results and Discussion}
\label{sec:3}

Excitation of the ZnO/(Zn,Mg)O MQW in the vicinity of the T$_\text{A}$ trion or the X$_\text{A}$ exciton results in a FWM signal in form of photon echoes. Figure \ref{transients}(a) displays a typical FWM transient measured at energy 3.378~eV after application of the sequence of three laser pulses with the delays $\uptau_{12}=26.7$~ps and $\uptau_{23}=13.3$~ps. It consists of three echo pulses occurring at their expected locations: The PE$_{12}$ at $\uptau_\text{Ref}= 2\uptau_{12}=53.4$~ps and PE$_{13}$ at $\uptau_\text{Ref}= 2(\uptau_{12}+\uptau_{23})=80$~ps are the primary photon echoes (PE) resulting from the 1-2 and 1-3 two-pulse sequences, respectively. The three-pulse (stimulated) photon echo (SPE) occurs at $\uptau_\text{Ref}= 2\uptau_{12}+\uptau_{23}\approx67$~ps. In a two-pulse echo experiment, variation of the $\uptau_{12}$ delay results in an exponential decay of the PE amplitude, from which the coherence time (irreversible dephasing time) $T_2$ of the optical excitation can be obtained. Figures \ref{transients}(b) and \ref{transients}(c) demonstrate PE signals detected on the T$_\text{A}$ trion at energy 3.378~eV and the X$_\text{A}$ exciton at energy 3.394~eV, respectively. From the amplitudes of these PE transients we evaluate the trion and the exciton coherence times to be $T_2=30$~ps and 4.5~ps for T$_\text{A}$ trion and X$_\text{A}$ exciton.

Fine tuning of the laser photon energy allows us to study the coherent dynamics of trions and excitons in the MQW in much greater detail. We perform two types of PE measurements with spectral resolution. In the first experiment, we fix the delay between the first and the second pulses, $\uptau_{12}$, and detect the PE amplitude at $\uptau_\text{Ref}= 2\uptau_{12}$ while scanning precisely the laser energy in the range 3.365--3.420~eV. In this way we measure spectrally resolved the PE amplitude for a certain delay. In the second experiment, we fix the excitation energy and detect the PE amplitude while scanning the $\uptau_{12}$ delay. This gives us the accurate PE kinetics at a certain transition energy.

\begin{figure}[t]
	\vspace{5mm}
	\includegraphics[width=\linewidth]{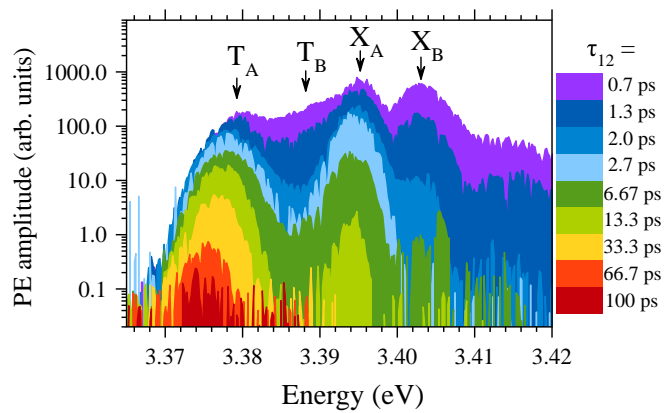}
	\caption{Two-pulse photon echo spectra measured from ZnO/(Zn,Mg)O MQW at different pulse delays $\tau_{12}$ indicated on the right hand side.}
	\label{spectra}
\end{figure}

PE spectra measured in the first experiment for a number of $\uptau_{12}$ delays are shown in Fig.~\ref{spectra}(a). The laser pulse energy is tuned and precisely detected by a spectrometer while the PE amplitude is measured at a certain $\uptau_{12}$ delay. Despite of the broadening by the laser pulse width (1.7~meV), the main spectral features are clearly observed. In these spectra we see not only the T$_\text{A}$ trion and the X$_\text{A}$ exciton giving echo signals at delays $\gtrsim$10~ps, but also the T$_\text{B}$ trion around 3.388~eV energy and X$_\text{B}$ exciton around 3.403~eV energy present only at short delays up to several ps. The energy positions of these transitions correspond well to those in the absorption spectrum [Fig.~\ref{setup}(a)]. The echo signal at energies $>3.41$~eV originates supposedly from texcited states of the QW excitons. From comparison of the spectra of PL and PE detected at $\uptau_{12}=1.3$~ps we prove the T$_\text{A}$ trion Stokes shift of 6~meV and evaluate the T$_\text{A}$ trion binding energy of 14~meV. The binding energy of the T$_\text{B}$ trion can be roughly estimated to be 15~meV. A comparable binding energy of the T$_\text{A}$ trion of 13~meV was measured earlier in a ZnO/(Zn,Mg)O MQW sample with similar composition \cite{PulsPRB2012}. We also observe a low-energy shift of the T$_\text{A}$ and X$_\text{A}$ spectral lines with increasing $\uptau_{12}$ delay.

\begin{figure*}[t]
	\vspace{5mm}
	\includegraphics[width=\linewidth]{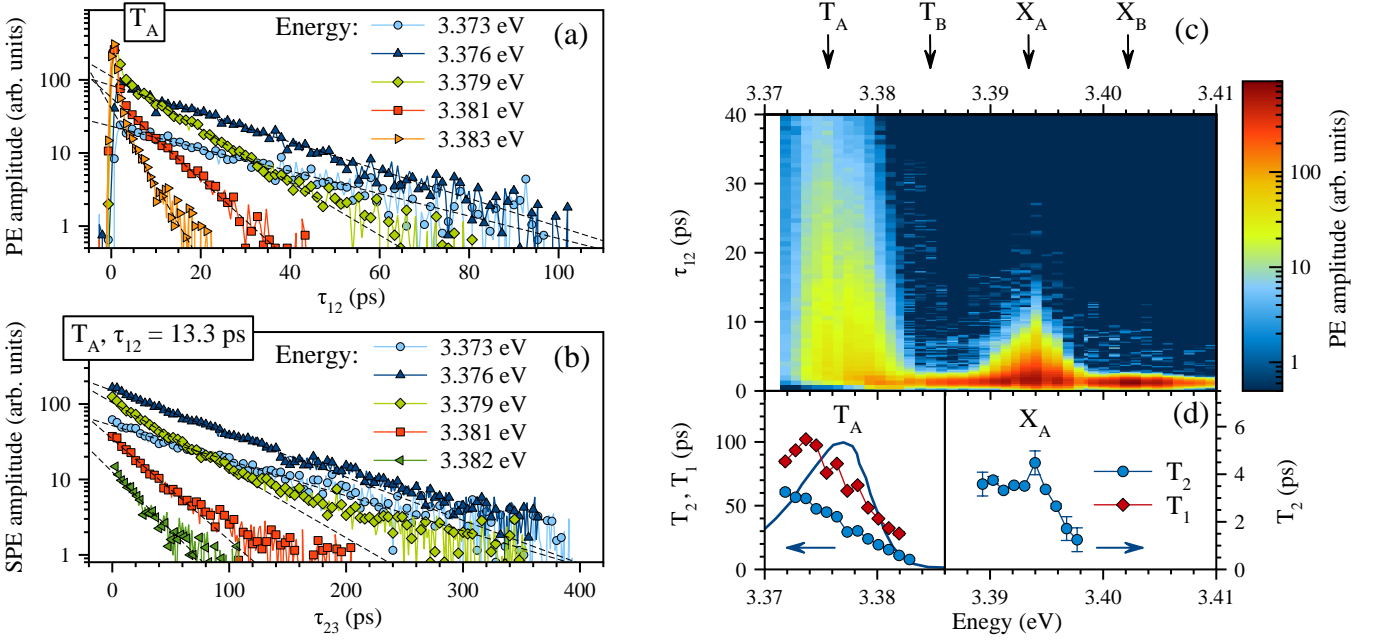}
	\caption{Two-pulse and three-pulse photon echo kinetics with spectral resolution at $T=1.5$~K: PE (a) and SPE (b) kinetics as functions of trion energy with single-exponential fittings (dashed line). The SPE is measured at $\uptau_{12}=13.3$~ps. (c) Intensity plot of the PE amplitude measured as a function of the pulse delay $\tau_{12}$ and laser energy. The spectral positions of the trions T$_\text{A}$, T$_\text{B}$ and excitons X$_\text{A}$, X$_\text{B}$ are indicated with markers. (d) Spectral dependencies of the coherence time $T_2$ and the population decay time $T_1$ extracted from the PE and SPE kinetics of the trion (left panel) and exciton (right panel). The PL spectrum is shown by the solid line on the left panel.}
	\label{PE_SPE}
\end{figure*}

In order to study the dynamics of the photon echoes we measure the temporal decays of the PE and SPE amplitudes by scanning the pulse delays $\uptau_{12}$ and $\uptau_{23}$, respectively. The results of these measurements accomplished with spectroscopic resolution are summarized in Fig.~\ref{PE_SPE}. The panoramic spectrum of the PE decays in Fig.~\ref{PE_SPE}(c) represents a set of PE kinetics measured with an energy step of about 1~meV in the vicinity of the A and B trions and excitons. Here, we clearly see the longer-lived PE signal in the range of the T$_\text{A}$ and X$_\text{A}$ features. The PE signal in the range of the T$_\text{B}$ and X$_\text{B}$ transitions is strong, but short-lived. The T$_\text{A}$ PE kinetics, shown in Fig.~\ref{PE_SPE}(a), follow mostly a single-exponential decay except for the kinetics measured at laser pulse energies $>3.376$~eV which demonstrate a sharp spike at the small delays, which corresponds to the contribution of the sub-ps T$_\text{B}$ trion kinetics. From fitting these kinetics by the exponential decay, we obtain the spectral dependencies of the A trion and exciton coherence times $T_2$, which are plotted in Fig.~\ref{PE_SPE}(d). For both transitions T$_\text{A}$ and X$_\text{A}$ we see monotonous increases of $T_2$ with decreasing energy. Trion coherence times up to $T_2=60$~ps are measured. The X$_\text{A}$ exciton coherence time reaches $T_2=4.5$~ps, which is substantially longer than that for the free exciton that was observed in an epitaxial layer showing phase relaxation on a sub-ps timescale\cite{PoltavtsevZnO2017}. The exciton coherence time can be shortened by efficient trion formation due the rather large resident electron density, which also results in the weak X$_\text{A}$ PL signal [not even seen in the linear intensity plot in Fig.~\ref{setup}(a)]. It is remarkable that in the QW both the T$_\text{B}$ trion and X$_\text{B}$ exciton exhibit such a short coherent dynamics ($T_2<1$~ps), being strongly different from that of the T$_\text{A}$ trion and X$_\text{A}$ exciton, respectively. We attribute this difference to the rapid energy relaxation from the higher energy B subband to the lower energy A subband of the trion and exciton transitions.

Using the three-pulse excitation protocol and detecting the SPE amplitude we can measure the population dynamics. In this experiment we fix the $\uptau_{12}=13.3$~ps delay and scan the $\uptau_{23}$ delay, while measuring the SPE amplitude at $\uptau_\text{Ref}= 2\uptau_{12}+\uptau_{23}$. The population dynamics detected at different energies within the T$_\text{A}$ trion line are shown in Fig.~\ref{PE_SPE}(b). These kinetics do not perfectly match single-exponential decays, but rather manifest a short component and a long tail. The short part can be satisfactorily fitted with the form $\sim\exp(-\uptau_{23}/T_1)$, which gives the spectral dependence of the population decay time $T_1$ displayed in Fig.~\ref{PE_SPE}(d). We see that the trion $T_1$ time grows monotonously up to 100~ps resulting approximately in a constant ratio $T_2/T_1\approx0.5$. From analysis of the long component of the SPE decay we extract population decay times up to 200--250~ps (not shown), which is in agreement with the T$_\text{A}$ trion radiative lifetime of 200~ps measured recently in the PL kinetics of a similar ZnO/(Zn,Mg)O MQW, using a streak-camera \cite{PulsPRB2016}.

\begin{figure}[t]
	\vspace{5mm}
	\includegraphics[width=\linewidth]{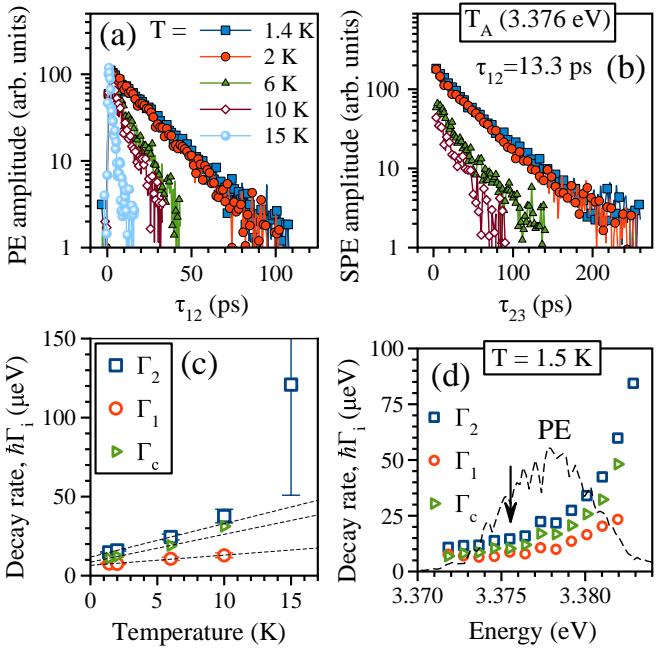}
	\caption{Trion coherence time $T_2$ at various excitation intensities and temperatures: Variation of the trion PE kinetics (a) and SPE kinetics at $\uptau_{12}=13.3$~ps (b) detected at energy 3.376~eV with temperature. (c) Temperature dependences of the phase relaxation rate $\Gamma_2$ (squares), population decay rate $\Gamma_1$ (circles) and pure dephasing rate $\Gamma_c$ (triangles), obtained from the PE and SPE decays. The dashed lines are linear fits. (d) Spectra of the decay rates $\Gamma_2$, $\Gamma_1$ and $\Gamma_c$ obtained from the spectral data in Fig.~\ref{PE_SPE}. The dashed line is the PE spectrum measured at $\uptau_{12}=6.67$~ps. The arrow indicates the spectral position, at which the temperature dependence was measured.}
	\label{temperature}
\end{figure}

In order to uncover the role of phonons in the decaying coherent dynamics, we accomplished temperature-dependent two-pulse and three-pulse echo decay measurements on the T$_\text{A}$ trion at energy 3.376~eV. The decoherence and population kinetics measured at different temperatures are shown in Figs.~\ref{temperature}(a) and \ref{temperature}(b), respectively. The temperature dependences of the decoherence rate $\Gamma_2=\hbar/T_2$, the population decay rate $\Gamma_1=\hbar/T_1$ and the pure dephasing rate $\Gamma_c=\Gamma_2-\Gamma_1/2$ are plotted in Fig.~\ref{temperature}(c). We see from these graphs that the population dynamics is weakly influenced by the temperature up to $T=10$~K. The decoherence, however, is caused not only by acoustic phonons, whose contribution can be evaluated by a thermal coefficient $\upgamma\approx2~\mu$eV$/$K assuming a linear variation with temperature, but there is also a temperature-independent decoherence characterized by $\hbar\Gamma_d\approx9~\mu$eV, so that $\Gamma_c=\Gamma_d+\upgamma T$. To compare with the donor-bound exciton D$^0$X$_\text{A}$ (line $I_9$ at 3.357~eV) studied recently in a ZnO epitaxial layer\cite{PoltavtsevZnO2017}, where only the temperature-dependent part with a similar contribution of acoustic phonons ($\upgamma\approx2~\mu$eV$/$K) was present, the QW trions experience an additional phase relaxation. In order to check the effect of trion localization on its pure dephasing rate we address again the spectroscopic data from Fig.~\ref{PE_SPE} and plot the spectral dependencies of the $\Gamma_2$, $\Gamma_1$ and $\Gamma_c$ decay rates in Fig.~\ref{temperature}(d). Taking into account that the scattering on acoustic phonons at $T=1.5$~K can be neglected ($\gamma T\approx3~\mu$eV), we see that the main contribution to the trion pure dephasing is the temperature-independent decoherence ($\Gamma_d$), which increases strongly with rising laser energy.

\begin{figure}[t]
	\vspace{5mm}
	\includegraphics[width=\linewidth]{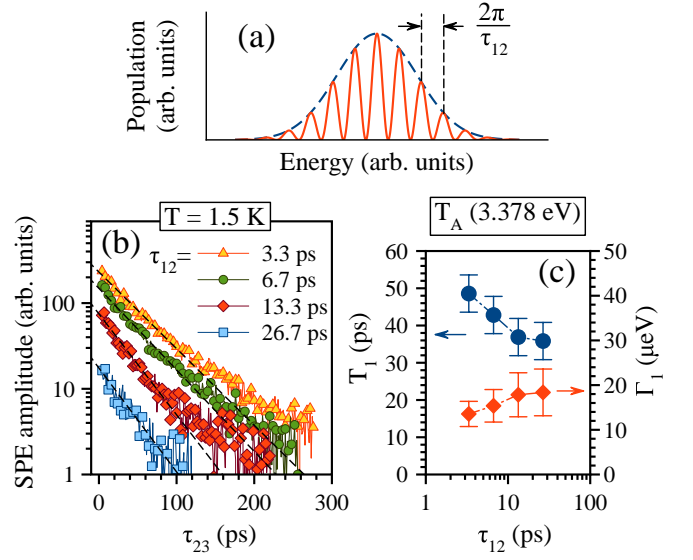}
	\caption{Trion population decay as a function of the $\uptau_{12}$ delay measured at energy 3.378~eV: (a) Schematic diagram with spectral population grating formed by the two exciting pulses delayed by $\uptau_{12}$. (b) SPE kinetics for different delays $\uptau_{12}$ (the points) with exponential fittings (the dashed lines). (c) Trion population decay times $T_1$ and population decay rate $\Gamma_1$ extracted from the SPE kinetics.}
	\label{spectral_diffusion}
\end{figure}

The observed low-energy shift of the A exciton and trion in the PE spectra measured at longer delays, as well as the monotonous increase of the trion $T_2$ and $T_1$ times with increasing energy can be explained by the localization of these particles on QW potential fluctuations. These fluctuations arise from random variations in QW thickness, in alloy composition, and in electric field due to an inhomogeneous distribution of the ionized donors in the doping layers. Additionally, because of the growth along the polar $c$-axis, there is also an inhomogeneous electric field due to the well width variations. The different particle localization degree might result in variations in the oscillator strength and, therefore, in the population decay rate. Since both the exciton and trion binding energies are clearly larger than the localization energy, the internal motion inside of exciton and trion are not strongly modified and the localization affects only the center-of-mass wave function. Respective model calculations were carried out so far only for the one-dimensional case of trions in a quantum wire \cite{EsserPSS2002}. For excitons localized in a QW, only a weak decrease of the oscillator strength is found for increasing localization \cite{Zimmermann1997}, that is much less dramatic than the $T_1$ change seen in Fig.~\ref{PE_SPE}(d). Therefore, we conclude that this change is dominantly the result of trion relaxation within the localization potential landscape. The observed strong growth of the temperature-independent trion decoherence rate $\Gamma_d$ with rising laser energy can be also explained by the trion localization. For a weakly localized trion the elastic scattering on the quantizing potential borders is the probable reason for the trion coherence loss.

We evaluate additionally the efficiency of spectral diffusion processes. These processes may arise in inhomogeneous ensembles from the excited state energy fluctuations due to various interactions with the environment. For precise quantification of spectral diffusion, different experimental methods were developed\cite{CundiffOSA2016}. Here we monitor it by means of the SPE decay measurements. Spectral diffusion may lead to a smoothing of the spectral population grating created by the first two exciting pulses as shown schematically in Fig.~\ref{spectral_diffusion}(a). The larger the $\uptau_{12}$ delay, the smaller the period of the spectral grating and, thus, the more sensitive the SPE decay to spectral diffusion processes. To check, we measure the SPE kinetics for different $\uptau_{12}$ delays as displayed in Fig.~\ref{spectral_diffusion}(b). They can be satisfactorily approximated by a single-exponent decay, from which we extract the population decay rate  $\Gamma_1(\uptau_{12})$, plotted in Fig.~\ref{spectral_diffusion}(c). From this plot we see indeed some increase of the population decay rate within $\hbar\Gamma_{SD}=5~\mu$eV at the $\uptau_{12}=26.7$~ps delay, which can be considered as approximate spectral diffusion rate. The effect of spectral diffusion on the trion dephasing, however, can be neglected ($\sim\hbar\Gamma_{SD}/2=2.5~\mu$eV). 

In line with our findings, the growth of the coherence time with increasing exciton localization measured by the photon echo technique was shown before on 9.6~nm wide GaAs/(Al,Ga)As QWs by Cundiff et al. \cite{CundiffPRB1992} In the studied QWs manifesting a Stokes shift of 1.2~meV, the phase relaxation rates varied in the range 0.01--0.11~ps$^{-1}$ ($\hbar/T_2=$7--72~$\mu$eV) at $T=5$~K. Also, excitons localized in the alloy fluctuations of CdS$_{0.4}$Se$_{0.6}$ crystals showed a similar variation of the coherence time across the exciton spectral line, resulting at $T=10$~K in an increase of $T_2$ from 260 up to 500~ps \cite{NollPRL1990}. Recently, however, a different spectral behavior of the phase relaxation rate of the donor-bound exciton in a CdTe/(Cd,Mg)Te single QW was observed\cite{PoltavtsevCdTe2017}. Although the heavy-hole trion demonstrated a monotonous increase of $T_2$ from 30~ps up to 100~ps, the D$^0$X transition located at just 0.6~meV lower energies showed a coherence time shortening from 100~ps down to 50~ps on its lower-energy side.

%We conclude here that, at low temperatures, two main processes affect the trion population dynamics. They include the energy relaxation from the higher energy trion states, i.e. spectral diffusion, and the trion localization. The variation in the trion phase relaxation rate over the energy spectrum, however, is mainly due to the changes in trion localization.

\section{Conclusions}
\label{sec:4}

Using spectrally narrow laser pulses we have studied the coherent dynamics of trions and excitons with different localization degree in a ZnO/(Zn,Mg)O MQW. We have been able to address the different localized states of the trion and to observe a strong reduction of the pure dephasing rate with increasing trion localization. Stronger trion localization is in favor of weaker many-body interactions, which are typically manifested through the effect of excitation-induced dephasing. In our study, however, the available optical excitation powers were insufficient to observed excitation-induced dephasing. It might be interesting to go beyond the $\upchi^{(3)}$ excitation regime and to drive the strongly localized trions with a higher optical field amplitude to obtain coherent Rabi flopping. In order to enhance the coupling between the trion and the driving optical field, ZnO/(Zn,Mg)O QWs could be placed in a Bragg \cite{SadofevAPL2011} or a Tamm-plasmon microcavity\cite{SalewskiPRB2017}. Moreover, strongly localized trions are interesting for further studies of spin-dependent optical coherent phenomena associated with the resident electron in their ground state. As resident electron spins in ZnO show a small $g$-factor dispersion \cite{BeschotenPSS2014}, photon echo-based effects exploited as optical memory on spins as realized recently for CdTe/(Cd,Mg)Te QWs\cite{LangerNP2014} appear to be quite promising for ZnO/(Zn,Mg)O QWs.

{\it Acknowledgments.} The authors appreciate financial support by the Deutsche Forschungsgemeinschaft (DFG) through the Collaborative Research Centre TRR 142 and the International Collaborative Research Centre 160. S.S. acknowledges DFG for the funding within SFB 951. I.A.S., S.V.P. and Yu.V.K. thank the Russian Foundation of Basic Research (RFBR) for partial financial support of their work (Contract No. 15-52-12016 NNIO\_a) and acknowledge St. Petersburg State University for the research grants 11.34.2.2012 and 11.42.664.2017. I.A.S. thanks RFBR for the research grant 18-32-00684 mol\_a.

%\bibliography{ZnO_QW}

\end{document}